\documentclass[aps,prd,preprint,superscriptaddress,showpacs,nofootinbib]{revtex4-1}

\usepackage{graphicx}
\usepackage{amssymb}

\begin{document}

\title{Energy loss as the origin of an universal scaling law of the elliptic flow }
\author{Carlota Andr\'es}
\email{carlota.andres@usc.es}
\affiliation{
        Instituto Galego de F\'\i sica de Altas Enerx\'\i as IGFAE, Universidade de Santiago de Compostela, E-15782 Santiago de Compostela (Galicia-Spain)}

\author{Mikhail Braun}
\email{mibraun@yandex.ru}
 \affiliation{Department of High-Energy Physics, Saint Petersburg State University, Russia}

\author{Carlos Pajares}
\email{pajares@fpaxp1.usc.es}
\affiliation{
        Instituto Galego de F\'\i sica de Altas Enerx\'\i as IGFAE, Universidade de Santiago de Compostela, E-15782 Santiago de Compostela (Galicia-Spain)}

\date{\today}

\def\beq{\begin{equation}}
\def\eeq{\end{equation}}

﻿

\begin{abstract}
  It is shown that the excellent scaling of the elliptic flow found for all centralities, species and energies from 
  RHIC to the LHC for $p_T$ less than the saturation momentum is a consequence of the energy lost by a parton 
  interacting with the color field produced in a nucleus-nucleus collision. In particular, the deduced shape of the 
  scaling curve describes correctly all the data. We discuss the possible extensions to higher $p_T$, proton-nucleus and proton-proton collisions as well as higher harmonics.
\end{abstract}

\maketitle

\section{Introduction}
The observed anisotropic flow \cite{b1,b2,b3} can exclusively be understood if the measured particles in the final state depend not only on the physical conditions realized locally at their production point, but also on the global symmetry of the event. This non-local information can solely emerge as a collective effect, requiring strong interaction among the relevant degrees of freedom, i.e., quarks and gluons. The study of higher
 harmonics has also shown very interesting features, including the ridge structure seen in A-A collisions \cite{b4,b5,b6,b7},  pPb collisions \cite{b8,b9} and even in high multiplicity pp collisions \cite{b10,b11,b12}.
The conventional understanding of the ridge is simply related to flow harmonics in a hydrodynamic scenario, where the description of the pPb ridge and especially the pp ridge is a challenge.
 The question is to what extent the ridge structure can be determined by the initial state effects and how these effects can be separated from the final state ones amenable to a hydrodynamic description. Along these lines, \cite{b13,b14}, it was pointed out that some scaling laws can be useful to disentangle initial state from final state effects . We have gone on this research showing that the elliptic flow for charged particles as well as for identified particles, including photons, satisfies a new scaling law \cite{b15}. This scaling cannot be derived from the geometrical scaling of the transverse momentum distributions \cite{b8,b9}. In this paper we go further in the understanding of the origin of this scaling showing that the interaction of the partons produced in the collision with the color field of the rest gives rise to this scaling. Moreover, we obtain the
detailed functional form of the scaling which shows a very good agreement with data.

\section{Universal scaling law}

 The universal scaling law proposed in reference \cite{b15} is
\beq
\frac{v_2(p_T)}{\epsilon Q^{A}_sL}=f(\tau).
\label{eq1}
\eeq

Here the eccentricity $\epsilon$ is defined by
\beq
\epsilon=\frac{2}{\pi}\int_0^{\pi/2}d\phi\cos 2\phi\frac{R^2-R_\phi^2}{R^2},
\eeq
where
\beq
R_\phi=\frac{R_A\sin(\phi-\alpha)}{\sin\phi},
\eeq
\beq
\alpha={\rm arcsin}\left(\frac{b}{2R_A}\sin\phi\right)
\eeq
and
\beq
R^2=<R_\phi^2>= \frac{2}{\pi}\int_0^{\pi/2}d\phi R_\phi^2.
\eeq

The scaling variable $\tau$ is
\beq
\tau=\left(\frac{p_T}{Q_s^A}\right)^2.
\eeq

$Q_s^A$ is the saturation momentum, $R_A$ is the radius of the nucleus and $L$
is the length
associated to the size of the collision area at a given impact parameter and energy. $Q_s^AL$ is the Knudsen number, i.e, the mean free path normalized to the length measured as the number of scattering centers. We take
\beq
  Q_s^A =Q_s^pA^{\alpha(s)/4}N_A^{1/12},
\eeq
where $N_A$ is the number of wounded nucleons, and $Q_s^p$ and $\alpha(s)$ are given
respectively
by
\beq
Q_s^p=Q_0\left(\frac{W}{p_T}\right)^{\lambda/2},\ \
\alpha(s)=\frac{1}{3}\left(1-\frac{1}{1+\ln(1+\sqrt{s/s_0})}\right).
\eeq

We take
\beq
Q_0=1\ {\rm GeV}, \ \ W=\sqrt{s} 10^{-3},\ \ \sqrt{s_0}=245\ {\rm GeV},\ \ \lambda=0.27.
\eeq

$L$ is taken as
\beq
L=\frac{1}{2}\left(1+N_A^{1/3}\right).
\eeq

The details and the motivation of this parametrization can be seen in reference \cite{b15} and references therein.

\begin{figure}
\includegraphics[width=\textwidth]{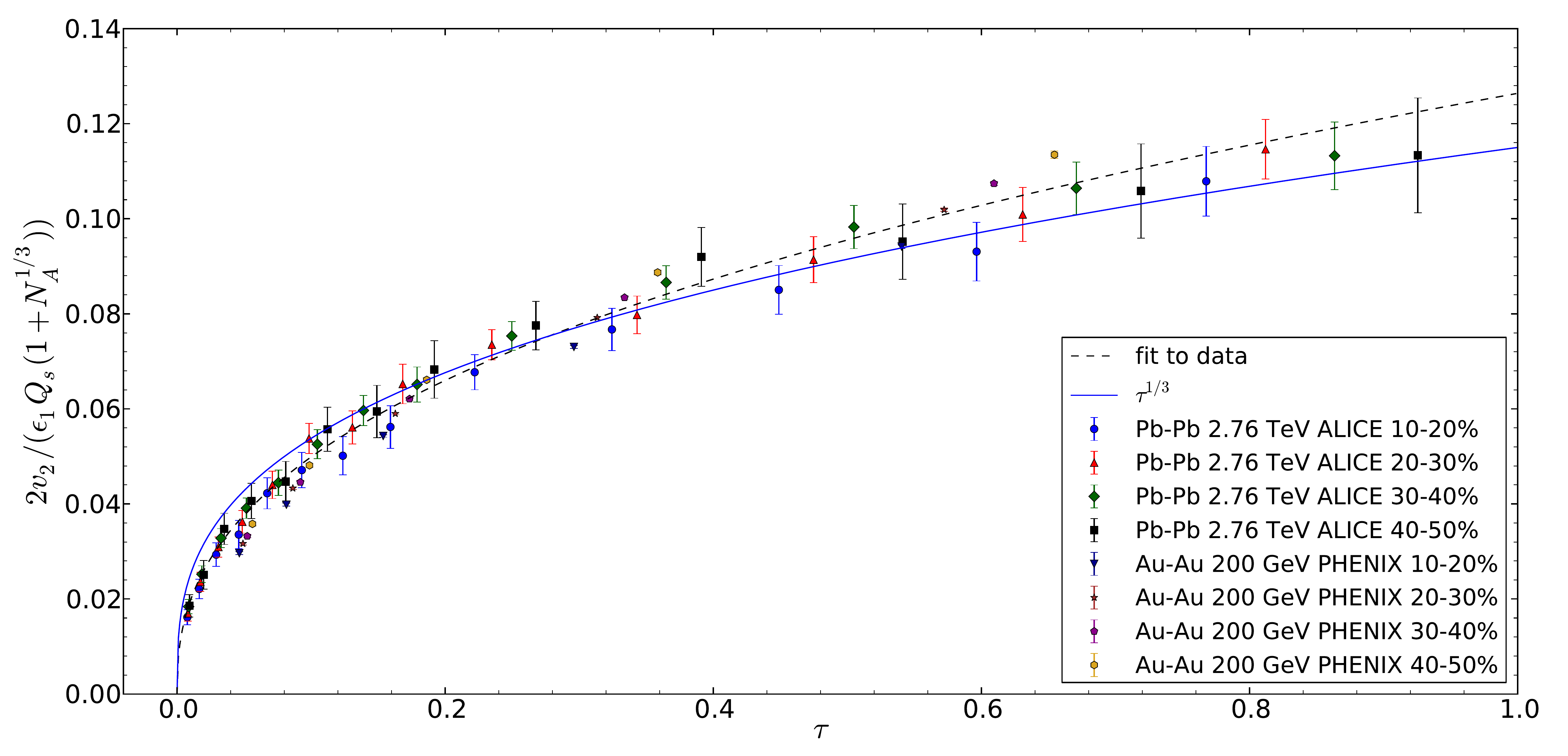}
\caption{(Color online.) $v_2$ divided by the product $\epsilon_1Q_s^AL$ for 10-20\%, 20-30\%, 30-40\% and 40-50\% Au-Au collisions at 200 GeV \cite{b15}, for 10-20\%, 20-30\%, 30-40\% and 40-50\% Pb-Pb collisions at 2.76 TeV \cite{b3} in terms of $\tau$. The dashed black line is a fit to data according to $a\tau^b$ with $a=0.1264\pm 0.0076$ and $b=0.404\pm 0.025$. The solid blue curve corresponds to $\tau^{1/3}$.}
\label{fig1}
\end{figure}

The experimental data for Au-Au at 200 GeV for the centrality range from 10\% to 50\% of PHENIX \cite{b15} and for PbPb at 2.76 TeV in the same centrality range \cite{b3} lie on the same curve that was fitted to the form $a\tau^b$  obtaining $a=0.1264\pm 0.0076$ and $b=0.404\pm 0.025$.
The fit is accurate for $p_T$ less than $Q_s$; see Fig.~\ref{fig1}.
The scaling law is also satisfied for pions, kaons and protons  \cite{b17,b18}. The photon data lie on the scaling curve too, although in this case the data present large errors bars due to large uncertainties \cite{b15}.

\section{Energy Loss}
    In a nucleus-nucleus collision strings are formed  among the partons of the colliding nucleons of both nuclei. In the transverse plane the strings can be seen as discs of small radius -- around 0.2 fm. As the energy or centrality of the collision increases, the number of strings increases and they start to overlap forming clusters of strings with a larger color field resulting from the vectorial sum of the individual
color fields of single strings. These clusters of strings decay similarly to a single string but with a larger string tension corresponding to their larger color field \cite{b21}. These decays roughly follow the Schwinger mechanism for producing pairs in the strong external field. The momentum distribution of these initial partons is azimuthally isotropic,
\beq
P(p_0)=Ce^{-p_0^2/\sigma},
\label{dist}
\eeq
where $p_0$ is the initial transverse momentum, $\sigma$ is the string tension and $C$ the normalization factor. It is important to point out that $p_0$ is different from the observed particle transverse momentum $p_{T}$,  because the parton has to pass through the cluster area emitting gluons on its way out. Therefore, in fact, the momentum distribution of the observed particles has the  the following form
\beq
P(p,\phi)=Ce^{-p_0^2(p,l(\phi))/\sigma},
\eeq
where $\phi$ is the azimuthal angle and $l(\phi)$ is the path length inside the nuclear overlap through which the observed particle has passed before
being observed.

Note that due to string tension fluctuations, distribution (\ref{dist}) is transformed into the thermal one
\beq
P(p_0)=Ce^{-p_0/T},
\label{dist1}
\eeq
where the temperature is $T=\sqrt{\sigma/2}$. In our calculation, this thermal distribution is used.

Radiative energy loss has been extensively studied for a parton passing through the nucleus or the quark-gluon plasma as a result of multiple collisions with the medium scattering centers \cite{b24,b25}. In our case, the situation is different: the created parton moves in the external gluon field of the string or cluster of strings, which, approximately, can be taken as constant and orthogonal to the direction of the parton. In the same vein as the mechanism of pair creation, one may assume that the reaction force due to radiation is similar to the QED case, where a charged particle is moving in an external electromagnetic field $E$. For an ultra-relativistic particle in a very strong field, this force causes an energy loss given by \cite{b26}
\beq
\frac{dp(l)}{dl}=-012e^2\left(eEp(l)\right)^{2/3},
\eeq
which leads to our quenching formula
\beq
p_0\left(p,l(\phi)\right)=p\left(1+\kappa p^{-1/3}T^{4/3}l(\phi)\right)^3,
\label{quench}
\eeq
where we have identified $eE/\pi=\sigma$ and introduced the dimensionless quenching coefficient $\kappa$. A fit to the experimental value of $v_2$ integrated over $p_T$ up to 4 Gev/c for Au-Au medium central collisions at RHIC has been done and this coefficient turned out to be small.

The possibility of using the QED formula for the QCD case may raise certain doubts. However, in \cite{b24} it was found by using the ADS-CFT correspondence that for the $N=4$ SUSY Yang Mills theory the energy loss of a colored charge moving in the external chromodynamic field is  essentially given by  the same formula as in QED case.

For small $\kappa$ we can approximate (\ref{quench}) as
\beq
p_0=p\left(1+\bar{\kappa} p^{-1/3}T^{4/3}l(\phi)\right),
\label{quench1}
\eeq
with $\bar{\kappa}=3\kappa$, so that the distribution in $p$ becomes
\beq
P(p,\phi)=Ce^{-p/T}e^{-\bar{\kappa}p^{2/3}T^{1/3}l(\phi)} .
\label{dist2}
\eeq

We expect the flow coefficient $v_2$ to be roughly proportional to the strength of the quenching -- it vanishes in absence of any quenching. On the other hand, it vanishes when quenching is isotropic in the azimuthal angle, which happens if the nuclear overlap is completely isotropic, i.e, in central collisions. Then, from Eq.(\ref{dist2}) we may expect
\beq
v_2 \sim p^{2/3}T^{1/3}\epsilon L,
\label{v21}
\eeq
where $\epsilon$ is the eccentricity of the nuclear overlap and $L$ is the path travelled by the particle inside the nucleus averaged over the azimuthal angles. To a good approximation $L$ is proportional to the average number of participants met by the particle on its path.
Note that $\epsilon$ and $L$ vary with the centrality in the opposite direction. At central collisions $\epsilon$ is small but $L$ attains its maximal value $R_A$. At peripheral collisions $\epsilon$ is large and $L$ is small. As a result, one expects a rather weak dependence on centrality; which has been confirmed by our previous calculations \cite{b27}.

Taking -- again roughly -- $T\sim Q_S^A$, we find from Eq.(\ref{v21}) that
\beq
\frac{v_2}{Q_s^A\epsilon L}\sim \left(\frac{p}{Q_s^A}\right)^{2/3}=\tau^{1/3}.
\label{scaling}
\eeq

In Fig.~\ref{fig1} the experimental data of PHENIX and ALICE are shown versus $\tau^{1/3}$. Also the best fit of the form  $\propto \tau^b$  is shown, which gives a value of $b$ of $b=0.404$, not very different from $1/3$. Taking into account the rather crude approximations in deriving our scaling formula (\ref{scaling}) we find this result quite remarkable. It confirms our assumptions about quenching of partons inside the nuclear overlap.

\section{Discussion}

The result obtained for the scaling of the elliptic flow indicates that the energy loss due to the interaction of the emitted parton with the color field of the strings is its natural explanation. This description can be extended to collisions of smaller sizes as p-A or pp collisions. From the scaling law of Eq.(\ref{eq1}), we have computed $v_2(p_T)$ for different impact parameters using the Gaussian form for the proton profile function ~\cite{b15}. The obtained values are slightly larger than the recently reported by CMS and ATLAS collaborations. Probably, the Gaussian form is not the proper profile function for the proton.

The scaling law $\propto\tau^{1/3}$ is found to be valid for $p_T<Q_s^A$. Notice that for central Pb-Pb collisions at 
the LHC, $Q_s^A$ is close to 4 GeV/c, consequently, the scaling holds for not so low values of $p_T$. At high $p_T$
jet quenching and $p_T$ suppression mechanisms enter into play
and one would not expect the dependence $v_2\propto p_T^{2/3}$ to be valid. In fact, the LHC data show that the transverse momentum dependence is proportional to $p_T^b$ with $b$ close to 1/2 \cite{b30, b31}. This suggests that the scaling form would change from $\tau^{1/3}$ to $\tau^{1/4}$,
which happens if at high transverse momenta quenching Eq.(\ref{quench}) changes into
\beq
p_0\left(p,l(\phi)\right)=p\left(1+\kappa p^{-1/2}T^{3/2}l(\phi)\right)^2.
\label{quench2}
\eeq

Note that from this equation one concludes that at very large distance $l$ quenching grows as $l^2$ in agreement with the results obtained in the framework of the perturbative QCD \cite{b24}. From Eq.(\ref{quench2}) at small $\kappa$ and not so large distances, on purely dimensional grounds, one obtains indeed the scaling of Eq.(\ref{eq1}) with $f(\tau)\propto \tau^{1/4}$.

Checking this behavior would indicate that the origin of elliptic flow is the same at low and high $p_T$, namely, the energy loss.

\begin{figure}
\includegraphics[width=\textwidth]{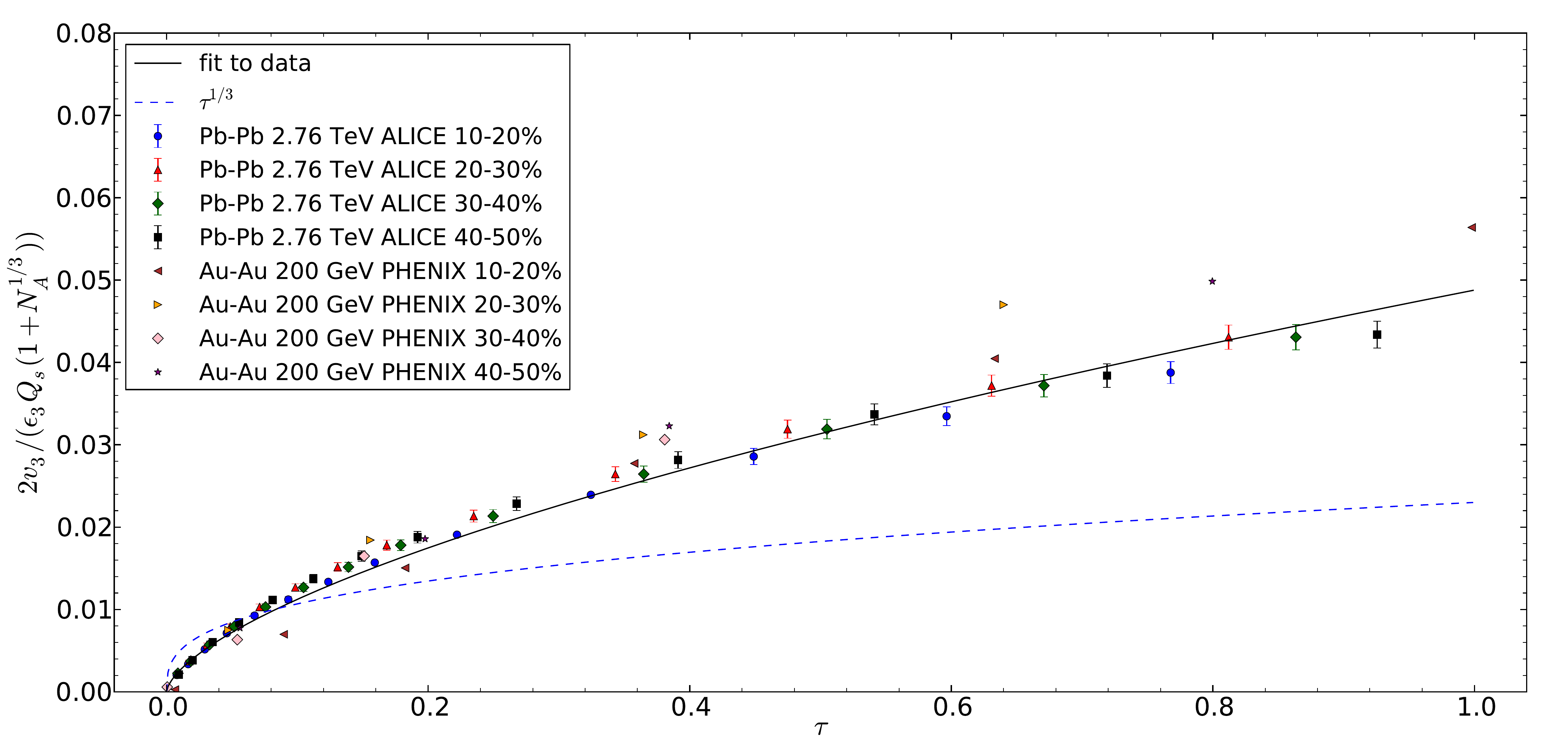}
\caption{(Color online.) $v_3$ divided by the product $\epsilon_3Q_s^AL$ for 10-20\%, 20-30\%, 30-40\% and 40-50\% Au-Au collisions at 200 GeV \cite{b32b}, for 10-20\%, 20-30\%, 30-40\% and 40-50\% Pb-Pb collisions at 2.76 TeV \cite{b32} versus $\tau$. The solid black line is a fit to data. The dashed blue curve corresponds to $\tau^{1/3}$.}
\label{fig2}
\end{figure}

Extension of this scaling to higher harmonics is questionable. It is known that $v_4$ and $v_5$ are not linear
with the  corresponding eccentricities contrary to the scaling in Eq.(\ref{eq1}). Both $v_3$ and $v_5$ are not purely 
geometrical and come from fluctuations, which implies some additional dynamics for their description. We have explored a possible scaling in $v_3$ in the simplest way, using eccentricity $\epsilon_3$ in Eq.(\ref{eq1}). In Fig.~\ref{fig2} we show the left hand side of Eq.(\ref{eq1}) as a function of $\tau$ using PHENIX and ALICE data for $v_3$ ~\cite{b32b,b32} and $\epsilon_3$ from ~\cite{b33}. In the latter reference multiplicity fluctuations described by a negative binomial distributions are included -- the parameter $k$ of these distributions, which determines fluctuations, is related to the nuclear profile function. 

We observe an approximate scaling, although its quality is not so good as for $v_2$. Also, $v_3$ does not rise as 
$\tau^{1/3}$, but considerably faster. This means that the energy loss alone cannot explain the scaling in $v_3$ and some additional dynamics, probably concerning the initial state, is necessary.

\section{Conclusions}

We have derived a universal scaling of the elliptic flow valid for all centralities and energies depending only on the ratio between the transverse and saturation momenta. We have also determined its concrete functional form $\propto\tau^{1/3}$, assuming that the energy loss of the parton emitted in A-A collisions and passing through the medium is given by the same expression as in QED. Comparison with RHIC and LHC data is very satisfactory.

We discuss possible extensions to smaller participants as pp or p-A collisions and to higher $p_T$, assuming that the energy loss mechanism is suitable in these cases.

Application to higher harmonics is also studied. In particular, it is shown that $v_3$ approximately satisfies a similar scaling although in this case the dependence on the scaling variable has a different functional form.

\section{Acknowledgments}
This work was supported by the project FPA2014-58243 C21P of Spain, by the Xunta de Galicia, by the grant RFFI 15-02-02097 of Russia  and the collaboration agreement between Saint-Petersburg and Santiago de Compostela Universities. C. Andr\'es thanks the Spanish Ministry of Education, Culture and Sports for financial support (grant FPU2013-03558).


\begin{thebibliography}{100}
%
\bibitem{b1} A. Adare et al. (PHENIX Collaboration), Nucl. Phys. {\bf A 757} 184 (2005).
%
\bibitem{b2} J. Adams et al. (STAR Collaboration), Nucl. Phys. {\bf A 757} 108 (2005).
%
\bibitem{b3} K. Aaamodt et al. (ALICE Collaboration), Phys. Rev. Lett. {\bf 105} 252302 (2010).
%
\bibitem{b4} J. Adams et al. (STAR Collaboration), Phys. Rev. {\bf C 78} 064917 (2006).
%
\bibitem{b5} A. Adare et al. (PHENIX Collaboration), Phys. Rev. {\bf C 78} 014961 (2006).
%
\bibitem{b6} B. Alver et al. (PHOBOS Collaboration), Phys. Rev. {\bf C 81} 034915 (2010).
%
\bibitem{b7} S. Chatrchyan et al. (CMS Collaboration), Eur. Phys. J, {\bf C 72} 2612 (2012).
%
\bibitem{b8} B. Abelev et al. (ALICE Collaboration), Phys. Lett. {\bf B 719} 24 (2013).
%
\bibitem{b9} S. Chatrchyan et al. (CMS Collaboration), Phys. Lett. {\bf B 718} 795 (2013).
%
\bibitem{b10} V. Khachatryan et al. (CMS Collaboration), IHEP {\bf 09} 091 (2016).
%
\bibitem{b11} V. Khachatryan et al. (CMS Collaboration), arXiv:1606.06148 [hep-ph].
%
\bibitem{b12} M. Aaboud et al. (ATLAS Collaboration), arXiv: 1606.08170 [hep-ex].
%
\bibitem{b13} R. Lacey et al., arXiv: 1105.3782 [nucl-ex].

\bibitem{b14} G. Torrieri et al., Phys. Rev.  {\bf C 84} 024908 (2014).
%
\bibitem{b15} C. Andr\'es, J. Dias de Deus, A. Moscoso, C. Pajares and C. Salgado, Phys. Rev. {\bf C 92} 034961 (2015).
%
\bibitem{b16} A. Adare et al. (PHENIX Collaboration), Phys. Rev. Lett. {\bf 98} 162301 (2007).
%
\bibitem{b17} S. S. Adler et al. (PHENIX Collaboration), Phys. Rev. {\bf C 69} 034909 (2004).
%
\bibitem{b18} B. Abelev et al. (ALICE Collaboration), Phys. Rev. {\bf 88} 044909 (2013).
%
\bibitem{b19} A .Adare et al. (PHENIX Collaboration), Phys. Rev. Lett. {\bf 109} 122302 (2012).
%
\bibitem{b20} D. Lohner (for ALICE Collaboration), J. Phys. Conf. Ser.{\bf 446} 012028 (2013).
%
\bibitem{b21} M. A. Braun et al., Phys. Rep. {\bf 599} 1-50 (2015).
%
\bibitem{b22} A. Bialas, Phys. Lett. {\bf B 466} 301 (1999).
%
\bibitem{b23} J. Dias de Deus and C. Pajares, Phys. Lett. {\bf B 642} 455 (2006).
%
\bibitem{b24} R. Baier et al., Nucl. Phys.  {\bf B 483} 291 (1997) {\bf B 484} 265 (1997).
%
\bibitem{b25} Y. Mehtar-Tani, C. A. Salgado and K. Tywoniuk, Phys. Rev. Lett. {\bf B 106} 122002 (2011).
%
\bibitem{b26} R. Baier, A. I. Nikishov and V. I. Ritus, Sov. Phys. Uspekhi {\bf 13} 303 (1970).
%
\bibitem{b27} M. A. Braun, C. Pajares and V. V. Vechernin, Nucl. Phys.  {\bf A 906} 14 (2013).
%
\bibitem{b28} M. A. Braun, C. Pajares and V. V. Vechernin, Eur. Phys. J. {\bf A 51} 44 (2015).
%
\bibitem{b29} A. Mikhailov, arXiv: 0305196 [hep-th].
%
\bibitem{b30} M. Spousta and B. Cole, Eur. Phys. J., {\bf C 76} 50 (2016).
%
\bibitem{b31} M. Spousta, arXiv: 1606.00903 [hep-ph].
%
\bibitem{b32b} A. Adare et al. (PHENIX Collaboration), Phys. Rev. Lett {\bf 107} 252301 (2011).
%
\bibitem{b32} K .Aaamodt et al. (ALICE Collaboration), Phys. Rev. Lett. {\bf 107} 032301 (2011).
%
\bibitem{b33} A. Dumitru and Y. Nara, Phys. Rev. {\bf C 86} 034906 (2012).
%
\end{thebibliography}
\end{document}